\documentclass[twocolumn,aps,prl,amsmath,amssymb]{revtex4-1}
\usepackage{graphicx}% Include figure files
\usepackage{dcolumn}% Align table columns on decimal point
\usepackage{bm}% bold math
\usepackage{multirow}

\usepackage{amssymb}
\usepackage{amsmath}
\usepackage{graphicx}
\usepackage{xcolor}
\usepackage[colorlinks,citecolor=blue,linkcolor=red,urlcolor=blue]{hyperref}
\usepackage{braket}
\usepackage{CJK}
\usepackage{indentfirst}
\usepackage{amsmath}
\usepackage{cases}

\begin{document}
\title{Driven Critical Dynamics in Tricitical Point}
\author{Ting-Long Wang$^{1}$}
\author{Yi-Fan Jiang$^{1}$}
\email{jiangyf2@shanghaitech.edu.cn}
\author{Shuai Yin$^{2,3}$}
\email{yinsh6@mail.sysu.edu.cn}

\affiliation{$^1$School of Physical Science and Technology, ShanghaiTech University, Shanghai 201210, China}
\affiliation{$^2$Guangdong Provincial Key Laboratory of Magnetoelectric Physics and Devices, School of Physics, Sun Yat-Sen University, Guangzhou 510275, China}
\affiliation{$^3$School of Physics, Sun Yat-Sen University, Guangzhou 510275, China}

\date{\today}
\begin{abstract}
The conventional Kibble-Zurek (KZ) mechanism, describing driven dynamics across critical points based on the adiabatic-impulse scenario (AIS), have attracted broad attentions. However, the driven dynamics in tricritical point with two independent relevant directions has not been adequately studied. Here, we employ time dependent variational principle to study the driven critical dynamics at a one-dimensional supersymmetric Ising tricritical point. For the relevant direction along the Ising critical line, the AIS apparently breaks down. Nevertheless, we find that the critical dynamics can still be described by the KZ scaling in which the driving rate has the dimension of $r=z+1/\nu_\mu$ with $z$ and $\nu_\mu$ being the dynamic exponent and correlation length exponent in this direction, respectively. For driven dynamics along other direction, the driving rate has the dimension $r=z+1/\nu_p$ with $\nu_p$ being the other correlation length exponent. Our work brings new fundamental perspective into the nonequilibrium critical dynamics near the tricritical point, which could be realized in programmable quantum processors in Rydberg atomic systems. 
\end{abstract}

\maketitle

{\bf Introduction}--- Nonequilibrium quantum critical dynamics represents a significant facet of quantum critical phenomena, attracting extensive investigations form both theoretical and experimental aspects. For the driven critical dynamics, the celebrated Kibble-Zurek (KZ) mechanism gives a unified description on the scaling of topological defects generated after the quench~\cite{Kibble1976,Zurek1985}. Crucial in the original KZ mechanism is the adiabatic-impulse scenario (AIS)~\cite{Dziarmaga2010review,Polkovnikov2011rmp}. By tuning a parameter $h$ with dimension $1/\nu$ to cross its critical value $h_c$ linearly with rate $R$, the evolution is separated into different stages. In the initial stage, the system evolves adiabatically as the correlation time $\tau$ is smaller than the transition time $t \sim |h-h_c|/R$. Usually, an energy gap $\Delta$ is required to ensure the adiabaticity as $\tau\sim \Delta^{-1}$. The adiabaticity ceases to hold inevitably near the critical point since $\Delta$ becomes small in the critical region and an impulse region then appears when $\tau>t$~\cite{Kibble1976,Zurek1985,Dziarmaga2010review,Polkovnikov2011rmp}. The boundary between these two regions dictates a frozen time~\cite{Dziarmaga2010review,Polkovnikov2011rmp}. The finite-time scaling (FTS) theory shows that this frozen time scale determines the typical time scale in the impulse region and controls dynamic scaling behaviors of macroscopic quantities~\cite{Zhifangxu2005prb,Gong2010njp,Yin2014prb,huangyy2014prb,Feng2016prb}. Both the KZ mechanism and the FTS have inspired a plethora of investigations~\cite{Zoller2005prl,Dziarmaga2005prl,PhysRevB.72.161201,Du2023,Ko2019,PhysRevLett.129.227001,sciadv.aba7292,science.abq6753,PhysRevB.78.144301,PhysRevB.92.064419,Deng2008epl,Chandran2012prb,Huse2012prl,Yin2017prl,Liuchengwei2014prb,Sandvik2015prl,Clark2016science,zeng2024FTSGNY,zeng2024susy,PhysRevLett.134.010409}. In particular, recently, they have been used in the rapidly developing quantum devices to prepare quantum states and probe quantum criticality~\cite{Keesling2019,king2023nature,Ebadi2021,garcia2024resolving,king2024computational,PhysRevB.106.L041109,PRXQuantum,science.abo6587}.

On the other hand, as the watershed of second- and first-order phase transitions, the tricritical points can display peculiar critical phenomena that have drawn extensive interest in both statistical physics and condensed-matter physics. For instance, the famous deconfined quantum criticality was recently found to be a tricritical point beyond the Landau-Ginzburg-Wilson paradigm~\cite{Senthil2004,Zhao2020prl,Chester2024prl,Li2022jhep,Takahashi2024}. In Dirac systems, the chiral tricritical point exhibits unusual critical properties beyond the quantum-classical mapping~\cite{Yin2018prl,Yin2020prbfiqcp}. A typical case is the tricritical point in the Ising model~\cite{PhysRevLett.52.1575,FRIEDAN198537,Zamolodchikov1986,PhysRevLett.120.206403,doi:10.1126/science.1248253}. Remarkably, it was shown that emergent space-time supersymmetry, characterizing the invariance of the effective action under exchanging the fermion and boson degrees of freedom, can appear therein. Moreover, as its generalization, quantum critical/tricritical points with emergent supersymmetry have been uncovered in various systems~\cite{doi:10.1126/science.1248253,Jian2015prl,Jian2017prl,Balents1998NodalLT,PhysRevLett.90.120402,Lee2007EmergenceOS,Ponte_2014,PhysRevLett.119.107202,PhysRevLett.115.166401,PhysRevLett.105.150605,PhysRevLett.114.090404, PhysRevB.87.165145,lzx.18sciadv,PhysRevLett.126.206801,Franz2019PRB,PhysRevB.100.075153}.

These intriguing equilibrium behaviors imply that tricritical points can exhibit more diverse nonequlibrium dynamic behaviors. However, previous studies about the dynamics near quantum multicritical points focused on a few integrable models~\cite{Divakaran_2009,PhysRevB.80.241109,Mukherjee_2010,Patra_2011}. Thus, the study of the driven dynamics associated with tricritical points has yet to reach maturity, leaving substantial room for further exploration. In addition, recently, it was proposed that the Ising tricritical point with emergent space-time supersymmetry can be realized in Rydberg atomic systems~\cite{PhysRevLett.133.223401}, wherein the criticality can be probed by the external driving~\cite{Keesling2019}. The requirements from both the theoretical and experimental aspects motivate us to explore the driven dynamics across the tricritical point.

In contrast to usual critical points, which possess only one relevant direction, the tricritical point has two independent relevant directions. One of these directions lies along the critical line. For driving along this direction to cross the triritical point, the AIS breaks down as a result of the absence of the energy gap in the initial stage~\cite{PhysRevB.78.144301,PhysRevB.92.064419,Deng2008epl,zeng2024FTSGNY,Zhengb1996prl}. Some questions arise as follows: How to describe the driven dynamics in different relevant directions? Whether the KZ scaling is still applicable along the Ising critical direction?
 
To answer these questions, in this paper, we explore the driven critical dynamics near the supersymmetric Ising tricritical point by changing both two relevant parameters. For changing the parameter along the Ising critical phase with gapless Majorana fermionic excitations to cross the tricritical point, the dynamic scaling properties of both the bosonic order parameter and the fermion correlation are explored. We shown that the driving rate has the dimension of $r_\mu=z+1/\nu_\mu$, in which $z$ is the dynamic exponent and $\nu_\mu$ is the correlation length exponent of the tricritical point in this direction. Accordingly, we confirm that the KZ scaling is still applicable albeit the breakdown of the AIS. We attribute the validity of the KZ scaling to the fact that the condition $z'<z+1/\nu_\mu$ is satisfied, in which $z'$ is the dynamic exponent of the gapless phase~\cite{zeng2024FTSGNY}. In addition, as comparison, we also explore the driven dynamics along other direction and confirm that the KZ scaling, featured by another rate exponent $r_p=z+1/\nu_p$, works. Possible experimental realization in programmable quantum processors in Rydberg atomic systems is discussed.

%Recently, it was proposed that the Ising tricritical point can be realized in the Rydberg atomic systems.

%In this work, for the first time, driven dynamics in susy tricritical point: gapless initial state, two relevant directions %whereas for small driving rate the usual finite-size scaling is restored. 

{\bf Model and static properties}--- The Hamiltonian $H$ of the model we considered is~\cite{doi:10.1126/science.1248253}
\begin{equation}
H=H_\sigma+H_\mu+H_{\sigma\mu},
\end{equation}
with
\begin{subequations}
\begin{align}
    &H_\sigma=J_\sigma \sum_i \sigma_i^z\sigma_{i+1}^z-h_\sigma \sum_i\sigma_i^x, \label{h2a} \\
    &H_\mu=J_\mu\sum_i ( \mu_{i,a}^z\mu_{i,b}^z+ \mu_{i,b}^z\mu_{i+1,a}^z)-h_\mu \sum_i (\mu_{i,a}^x+\mu_{i,b}^x), \label{h2b} \\ &H_{\sigma\mu}=g\sum_i(\sigma_i^x\mu_{i,a}^z-\sigma_i^z\sigma_{i+1}^z\mu_{i,b}^z), \label{h2c}
\end{align}
\label{h2}
\end{subequations}
in which $\sigma^{x,z}$ and $\mu^{x,z}$ are Pauli matrix in $x$ and $z$ directions, $J_\sigma$ and $J_\mu$ are the coupling coefficients, $h_\sigma$ and $h_\mu$ are the strength of the transverse fields, and $g$ is the coupling between two spin chains. In the following, $J_\sigma$ and $J_\mu$ are chosen as unity. After the Jordan-Wigner transformation on $\sigma$-spin, the Hamiltonian $H$ can convert to~\cite{doi:10.1126/science.1248253} 
\begin{eqnarray}
   H&=&-i\sum_j[1-g\mu_{j+1/2}^z+(-1)^j(h_\sigma-1)]\chi_j\chi_{j+1} \nonumber \\
   &+&\sum_j \mu_{j-1/2}^z\mu_{j+1/2}^z-h_\mu \sum_j \mu_{j+1/2}^x, \label{h3}
\end{eqnarray}
in which $\chi$ represents the Majorana fermion operator. 

It was shown that a tricritical point appears at $h_{\sigma,c}=1$ and $h_{\mu,c}\approx 1.27$ for $g=0.5$ (see Supplemental materials). At the tricritical point, the supersymmetry emerges featured by identical anomalous dimensions for both the bosonic order parameter field $\mu^z$ and the fermionic operator $\chi$, namely $\eta_b=\eta_f=2/5$~\cite{doi:10.1126/science.1248253}. 

\begin{figure}[tbp]
\centering
  \includegraphics[width=\linewidth,clip]{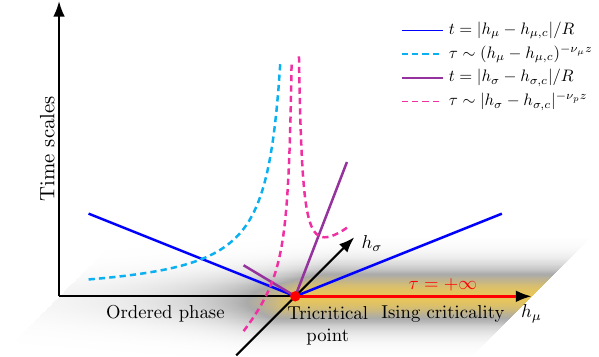}
  \vskip-3mm
  \caption{Sketch of the phase diagram near tricritical point of the Ising model and the protocol for driven dynamics. Around the tricritical dimension, there are two independent relevant directions. One is the $h_\mu$-direction for $h_\sigma=1$. Although the conventional AIS breaks down for driven dynamics from the Ising critical phase along this direct, we show that the dynamic scaling behaviors satisfy the FTS form in which the driving rate $R$ has the dimension of $r_\mu=z+1/\nu_\mu$. The driven dynamics along other relevant direction is explored by tuning $h_\sigma$ to cross the tricritical point, which exhibits a distinct FTS form in which $R$ has the dimension of $r_p=z+1/\nu_p$.}
  \label{figure1}
\end{figure}

There are two independent relevant directions around the tricritical point. One is the $h_\mu$-direction with fixed $h_\sigma=1$. Along this direction, for large $h_\mu$, the order parameter, namely the expectation value of $\mu^z$, is zero. In this way, the system is in the Ising critical phase and the fermion excitations are gapless. For small $h_\mu$, $\langle \mu^z\rangle$ becomes finite due to the  spontaneous symmetry breaking. Accordingly, the Majorana fermion gains a mass via the term $g\mu^z$. The correlation length exponent in the $h_\mu$-direction is $\nu_\mu=5/4$. 
In the perpendicular direction, namely $p$-direction, the correlation length exponent is $\nu_p=5/9$. In practice, to exactly determine this direction is very subtle. However, it was shown that for any arbitrary direction that has nonzero component in $p$-direction, the critical behavior is dominated by the variable in $p$-direction, since $1/\nu_p>1/\nu_\mu$~\cite{PhysRevE.92.022134}.

In the following, we utilize the time-dependent variational principle~\cite{TDVP} (TDVP) with matrix product state (MPS) ansatz to simulate the driven dynamics. The bond dimension of MPS up to $D=1000$ and time step $\tau=0.02\sim 0.04$ are taken to ensure the accuracy of the TDVP simulation. Although the original KZM concentrates on the scaling of the dependence of topological defects on the driving rate~\cite{Kibble1976,Zurek1985,Dziarmaga2010review,Polkovnikov2011rmp}, the FTS generalizes the scaling theory to more general quantities in the whole critical region. In this paper, we focus on the dynamic scaling behaviors exactly at the tricritical point for driven dynamics along different directions.

%Two chain: $\sigma$ chain comes from majorana fermion; $\mu$-chain is the boson field. At the tricritical point.

{\bf Driven dynamics along $h_\mu$-direction}--- We begin with the driven dynamics for fixing $h_\sigma=1$ and decreasing $h_\mu$ as $h_\mu=h_{\mu 0}-Rt$ with $h_{\mu 0}=1.8$. The corresponding initial state is in the Ising critical phase that hosts gapless Majorana excitations. Therefore, the correlation time in the initial stage is divergent and always larger that the transition time, implying the breakdown of the AIS in the original KZ dynamics.

\begin{figure}[btp]
    \centering  
    \includegraphics[width=1.0\linewidth]{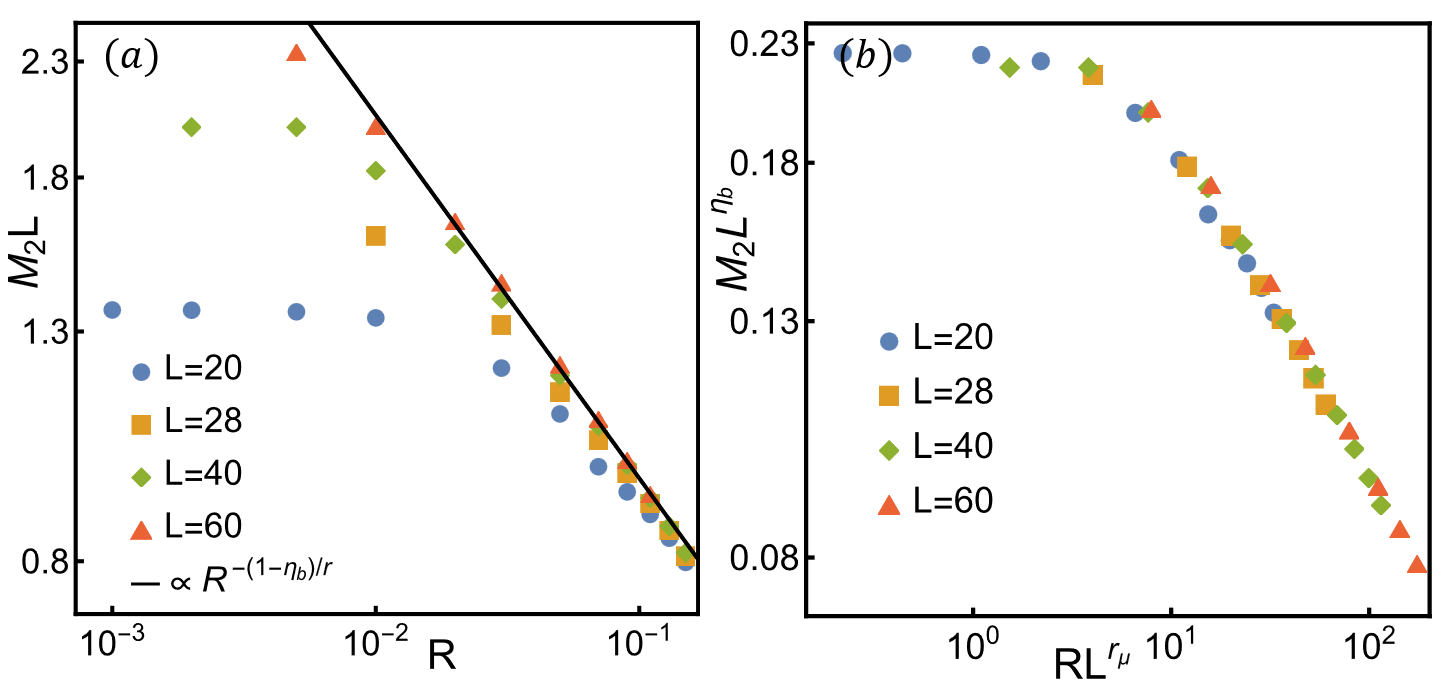}
    \caption{Scaling behavior of order parameter $M_2$ at the tricritical point when the system is driven from the gapless Ising critical phase by decreasing $h_{\mu}$. The solid line in (a) shows the power law $M_2\propto R^{-(1-\eta_b)/r_\mu}$ fitted to the $L=60$ data in large $R$ region. The rescaled curves in (b) collapse onto each other.}
    \label{Fig1}
\end{figure}

We firstly study the dynamics of the square of the order parameter defined as $M_2\equiv 1/L\sum_i\langle \mu_0^z\mu_i^z\rangle$ for large $R$. Fig.~\ref{Fig1} shows the dependence of $M_2$ on $R$ at the tricritical point. From Fig.~\ref{Fig1} (a), where we can find that $M_2\propto L^{-1}$ for a fixed $R$. Moreover, by comparing with a curve of the power function with exponent $(1-\eta_b)/r_\mu$, in which $\eta_b=2/5$ and $r_\mu=z+1/\nu_\mu$, we find the scaling relation of $M_2$
\begin{equation}
M_2(R,L)\propto L^{-1}R^{-(1-\eta_b)/r_\mu}.
\label{m21}
\end{equation}
Physically, Eq.~(\ref{m21}) reflects the memory of the initial state in the universal dynamics of the tricritical point under external driving. First, $M_2\propto L^{-1}$ stems from the property of $M_2$ for the initial state, since the $\mu^z$-spin in $H_\mu$-chain is short-range correlated in the large $h_\mu$ region. 
Second, the critical exponents of the tricritical point enter Eq.~(\ref{m21}) via the driving rate $R$. Since $M_2$ has the dimension of $\eta_b$, Eq.~(\ref{m21}) demonstrate that $R$ has the dimension of $r_\mu$, which is consistent with the KZ scaling. 

Accordingly, for a general $R$, the dynamic scaling of $M_2$ should obey
\begin{equation}
  M_2(R,L)=L^{-1}R^{-(1-\eta_b)/r_\mu} \mathcal{F}(RL^{r_u}),
  \label{m22}
\end{equation}
in which $\mathcal{F}$ is a scaling function. Note that Eq.~(\ref{m22}) is just the FTS form in usual critical point for driven dynamics from disordered phase~\cite{huangyy2014prb,Liuchengwei2014prb}. In addition, Eq.~(\ref{m22}) can convert to $M_2(R,L)=L^{-\eta_b} \mathcal{F}_1(RL^{r_u})$, from which the usual finite-size scaling of $M_2(R,L)\propto L^{-\eta_b}$ at the tricritical point can be recovered~\cite{PhysRevLett.119.107202}. After rescaling $M_2$ and $R$ as $M_2L^{\eta_b}$ and $RL^{r_\mu}$, respectively, we find that the rescaled curves collapse onto each other as shown in Fig.~\ref{Fig1}, confirming the scaling behavior in Eq.~(\ref{m22}). These results demonstrate that the KZ scaling still hold albeit the breakdown of AIS for gapless initial state.

\begin{figure}[btp]
    \centering  
    \includegraphics[width=1.0\linewidth]{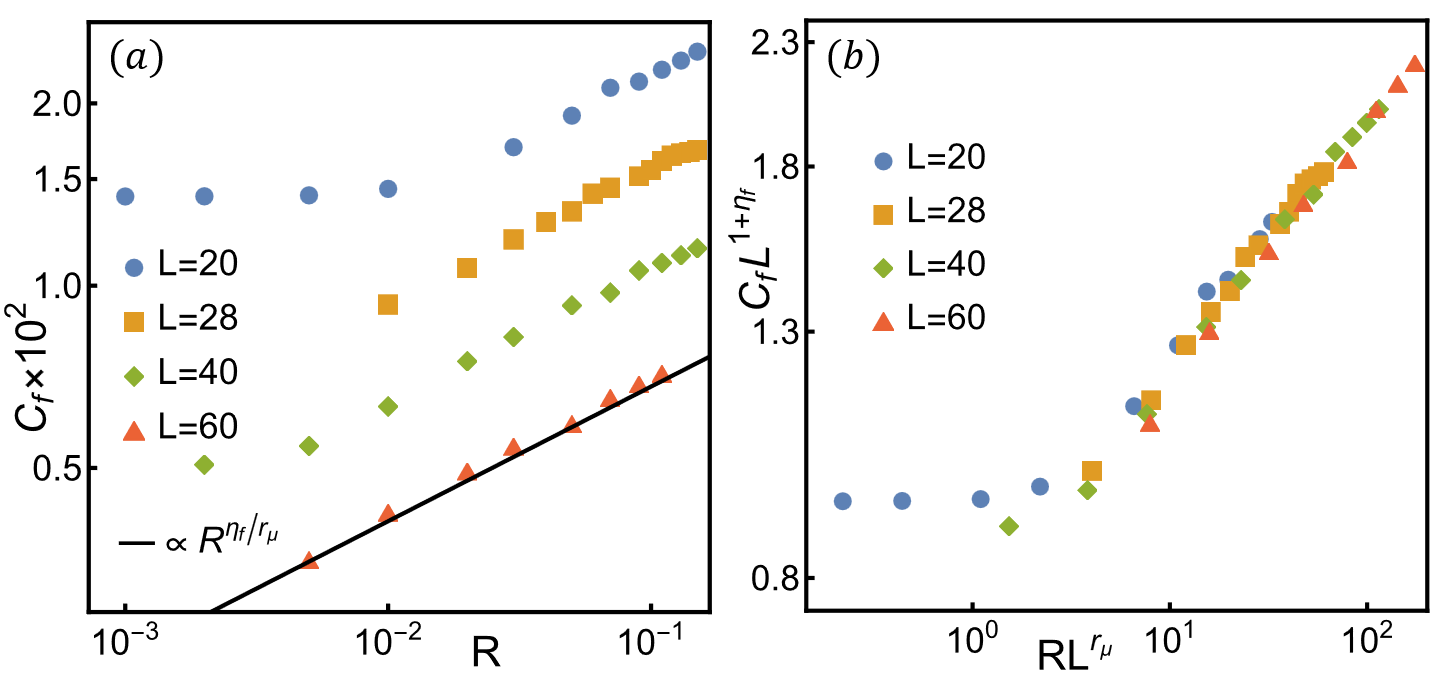}
    \caption{Scaling behavior of the half-chain fermion correlation $C_f$ at the tricritical point when the system is driven from the gapless Ising critical phase by decreasing $h_{\mu}$. The solid line in (a) shows the power law $C_f\propto R^{\eta_f/r_\mu}$ fitted to the $L=60$ data in large $R$ region. The rescaled curves in (b) collapse onto each other.}
    \label{Fig2}
\end{figure}

To further validate this conclusion, we then study the dynamic scaling of the fermion correlation defined as $C_f\equiv 1/L \sum_i \langle \chi_i \chi_{i+L/2} \rangle$ at the tricritical point. From Fig.~\ref{Fig2} (a), we find that for large $R$, the scaling relation of $C_f$ on $R$ is
\begin{equation}
C_f(R,L)\propto L^{-1}R^{\eta_f/r_\mu}.
\label{cf1}
\end{equation}
(See Supplemental materials). Similarly, Eq.~(\ref{cf1}) also combines the initial information, which is reflected in $C_f\propto L^{-1}$ stemming from the scaling of gapless Majorana excitations, with tricritical properties manifested in the exponent of $R$. Moreover, since $C_f$ has the dimension of $1+\eta_f$, Eq.~(\ref{cf1}) demonstrate that $R$ has the dimension of $r_\mu$, confirming the validity of the KZ scaling. Accordingly, the FTS form of $C_f$ should satisfy
\begin{equation}
  C_f(R,L)=L^{-1}R^{\eta_f/r_u} \mathcal{F}_2(RL^{r_u}), \label{cf2}
\end{equation}
which can be modified to $C_f(R,L)=L^{-1-\eta_f} \mathcal{F}_3(RL^{r_u})$, restoring the finite-size scaling $C_f(R,L)\propto L^{-1-\eta_f}$ for small $R$~\cite{PhysRevLett.119.107202}. This FTS form is confirmed in the scaling collapse of the rescaled curves of $C_fL^{1+\eta_f}$ versus $RL^{r_\mu}$ shown in Fig.~\ref{Fig2} (b).

In the equilibrium study of the tricritical point, the scaling behavior of the entanglement entropy is an important indicator to diagnose the critical properties. In particular, the half-chain entanglement entropy $S$ satisfies $S=(c/3)\log L$ in the critical phase/point with $c$ being the central charge. In the Ising critical phase, $c=c_I=1/2$; while at the tricritical point, $c=c_T=7/10$~\cite{doi:10.1126/science.1248253}. For the KZ dynamics with AIS in usual critical point, scaling of entanglement entropy was also studied~\cite{PhysRevA.75.052321}. Incorporating the scaling dimension of $R$ derived above, one can obtain the FTS form of $S$ at the tricritical point for driven dynamics from the Ising critical phase as follows
\begin{equation}
  S(R,L)=\frac{c_{T}}{3} \log L+ \mathcal{F}_4(RL^{r_\mu}).
  \label{s1}
\end{equation}
For large $R$, $S$ should retain the memory of the initial Ising critical phase. This requires scaling function $\mathcal{F}_4$ to obey $\mathcal{F}_4(RL^{r_\mu})=[(c_I-c_T)/3r_\mu] \log (RL^{r_\mu})$ such that Eq.~(\ref{s1}) tends to
\begin{equation}
S(R,L)=\frac{c_{I}}{3} \log L+\frac{(c_I-c_T)}{3r_u}\log{R}.
\label{s2}
\end{equation}

\begin{figure}[btp]
    \centering  
    \includegraphics[width=1.0\linewidth]{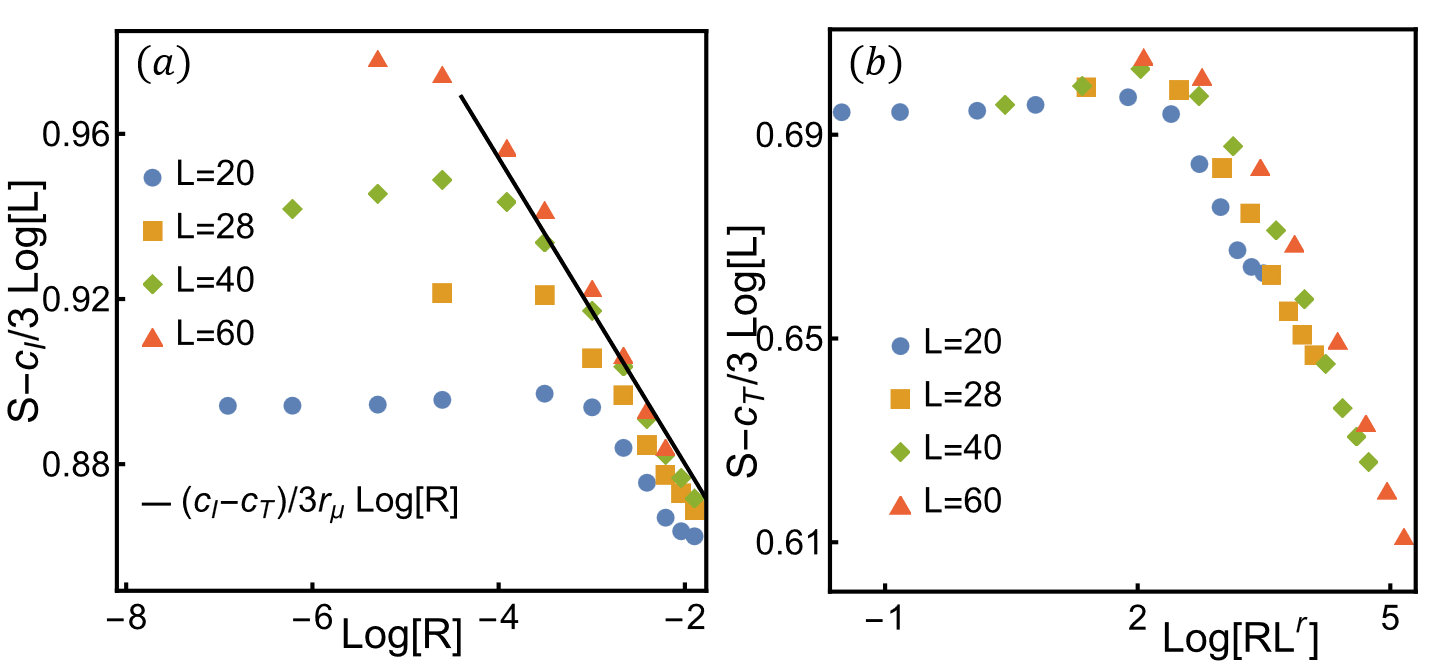}
    \caption{Scaling behavior of the half-chain entanglement entropy $S$ at the tricritical point when the system is driven from the gapless Ising critical phase by decreasing $h_{\mu}$. The solid line in (a) shows the logarithmic law $S-(c_I/3)\propto [(c_I-c_T)/3r_\mu] \log R$ fitted to the $L=60$ data in large $R$ region. The rescaled curves in (b) collapse onto each other.}
    \label{Fig3}
\end{figure}

Fig.~\ref{Fig3} (a) shows the curves of $S-(c_I/3)\log L$ versus $\log R$. In large $R$ region, the curve for large $L$ can be described by a line with the slope close to $(c_I-c_T)/3r_u$, consistent with Eq.~(\ref{s2}). The FTS form of $S$ in Eq.~(\ref{s1}) is confirmed by the collapse of the rescaled curves of $S-(c_T/3)\log L$ versus $RL^{r_\mu}$ illustrated in Fig.~\ref{Fig3} (b).

The above results reveal that although the AIS breaks down for driven dynamics from the Ising critical phase, the KZ scaling remains valid. To explore the reason, we inspect the Eqs.~(\ref{h2}) and (\ref{h3}) with $h_\sigma=1$. When the system is driven from critical phase $h_\mu>h_{\mu,c}$, $\langle \mu^z \rangle$ keeps zero in the initial stage. Thus, at mean-field level, the eigenstates of Eq.~(\ref{h2a}) or the gapless Majorana fermion part in Eq.~(\ref{h3}) remain intact. Consequently, the excitations in the initial stage are absent. In fact, it was shown that if the contributions beyond the mean field are taken into account, only the velocity of the Majorana fermion changes, while the linear dissipation of the gapless Majorana fermion keeps invariant~\cite{doi:10.1126/science.aao2934}. In this case, the density of excitation in the gapless phase obeys $n_{ex}\propto R^{1/z'}$ with $z'=1$ being the dynamic exponent of the Ising critical phase~\cite{Polkovnikov2008natphy}. In contrast, the KZ scaling of $n_{ex}$ for driving across the tricritical point is $n_{ex}\propto R^{1/r_\mu}$~\cite{Zoller2005prl,Dziarmaga2005prl,PhysRevB.72.161201}. Since the exponents $1/z'>1/r_\mu$, for the KZ dynamics with $R<1$, the critical properties near the tricritical point, rather than the critical phase, dominate. Note that recently a similar criterion in usual critical point for the validity of the KZ scaling with a gapless initial state was proposed~\cite{zeng2024FTSGNY}, which gives a unified explanation on the critical-point-dominated scaling behaviors reported previously~\cite{PhysRevB.78.144301,PhysRevB.92.064419,Deng2008epl}. Here, we successfully generalize the criterion to the tricritical point.

\begin{figure}[btp]
    \centering  
    \includegraphics[width=1.0\linewidth]{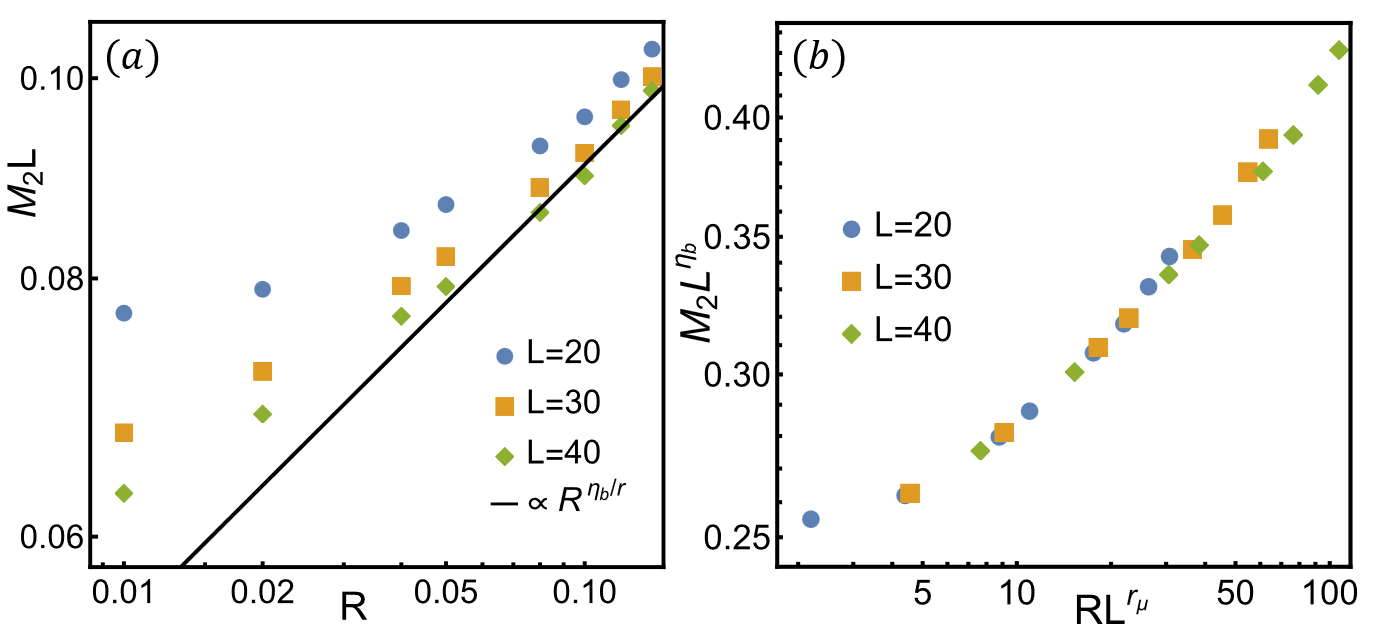}
    \caption{Scaling behavior of $M_2$ at the tricritical point for increasing $h_{\mu}$. The solid line in (a) shows the logarithmic law $M_2\propto R^{\eta_b/r_\mu}$ fitted to the $L=40$ data in large $R$ region. The rescaled curves in (b) collapse onto each other.}
    \label{Fig4}
\end{figure}

As a contrast, we then explore the driven dynamics by fixing $h_\sigma=1$ and increasing $h_\mu$ as $h_\mu=h_{\mu0}+Rt$ from the initial state $h_{\mu0}=0.6$ in the ordered phase with gapped fermion excitation. The driven dynamics near the tricritical point is similar to the case in usual critical point with an ordered initial state. Thus, the order parameter $M_2$ at the triciritical point should conform to
\begin{equation}
  M_2(R,L)=R^{\eta_b/r_u} \mathcal{F}_5(RL^{r_u}),
  \label{m23}
\end{equation}
where the memory from the ordered initial state, in which $M_2$ is almost independent of $L$, makes $M_2(R,L)\propto R^{\eta_b/r_u}$ for large $R$~\cite{Zhifangxu2005prb,huangyy2014prb}. This is verified in Fig.~\ref{Fig4} (a). In addition, Fig.~\ref{Fig4} (b) shows a nice collapse of the rescaled curves as expected according to Eq.~(\ref{m23}).

{\bf Driven dynamics along $h_\sigma$-direction}--- Different from usual critical point, the tricritical point possess two independent relevant directions. Although the definite one with $\nu_\mu=5/4$ is along the critical line, determining its perpendicular counterpart $p$-direction with $\nu_p=5/9$ seems quite subtle. For driven dynamics along an arbitrary dimension, which can be decomposed into a component in $h_\mu$-direction and a component in $p$-direction, the driving rate $R$ can consequently have two dimensions, namely $r_\mu=z+1/\nu_\mu$ and $r_p=z+1/\nu_p$. Since $r_p>r_\mu$, the driven dynamics should be dominated by the $p$-direction.

\begin{figure}[btp]
    \centering  
    \includegraphics[width=1.0\linewidth]{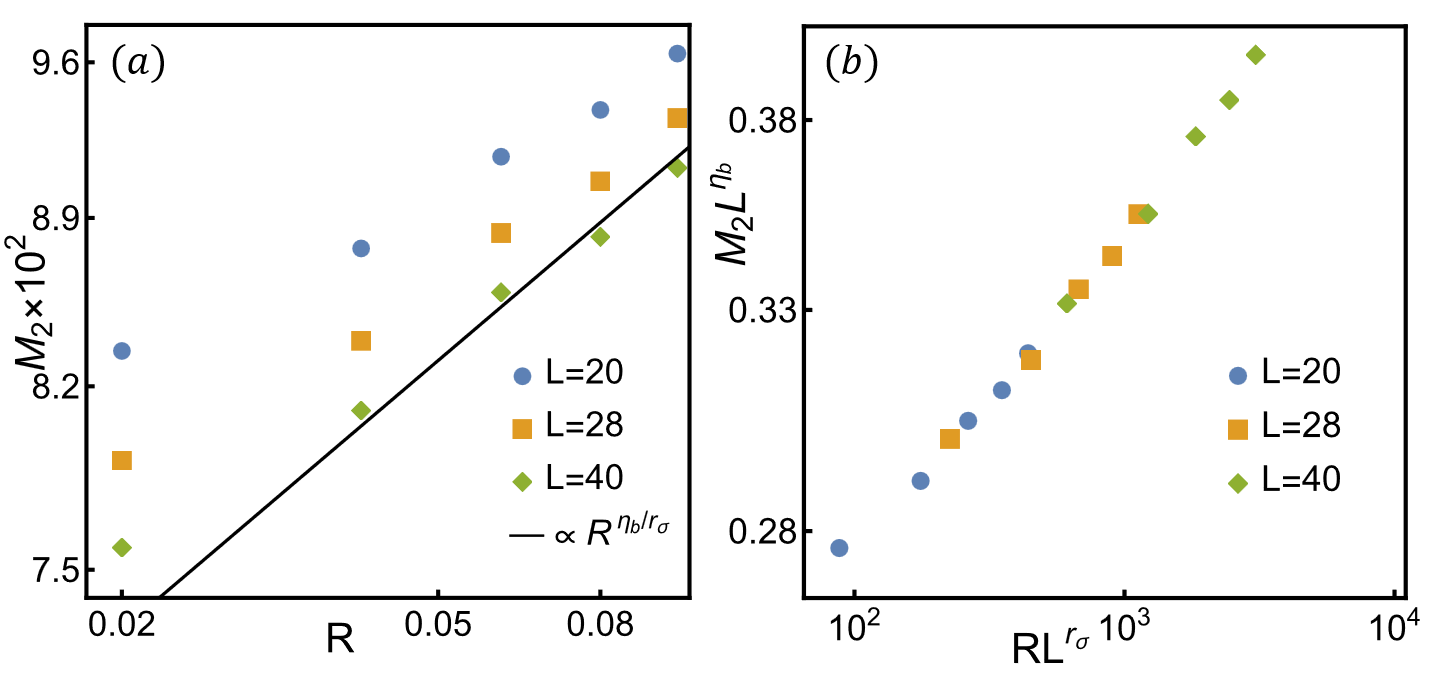}
    \caption{Scaling behavior of $M_2$ at the tricritical point for increasing $h_{\sigma}$. The solid line in (a) shows the logarithmic law $M_2\propto R^{\eta_b/r_\sigma}$ with $r_\sigma=r_p$ fitted to the $L=40$ data in large $R$ region. The rescaled curves in (b) collapse onto each other.}
    \label{Fig5}
\end{figure}

We take the driven dynamics along the $h_\sigma$-direction with $h_\mu$ fixed at $h_{\mu,c}$ as an example. Fig.~\ref{Fig5} shows the dependence of $M_2$ on $R$ at the tricritical point for a series of system sizes. For large $R$, we find from Fig.~\ref{Fig5} (a) that 
\begin{equation}
M_2(R,L)\propto R^{\eta_b/r_\sigma},
\label{m24}
\end{equation}
with $r_\sigma=r_p$ for large system size. Scaling analyses based on the dimension of $M_2$ suggest that the dimension of $R$ is $r_p$, confirming the discussion above.

The conclusion can be further verified by the inference for the full FTS form of $M_2$
\begin{equation}
  M_2(R,L)=R^{\eta_b/r_\sigma} \mathcal{F}_6(RL^{r_\sigma}).
  \label{m25}
\end{equation}
The scaling collapse of rescaled curves of $M_2$ versus $R$ for different $L$ shown in Fig.~\ref{Fig5} (b) confirms Eq.~(\ref{m25}), demonstrating that the typical exponent along $h_\sigma$-direction is $r_\sigma=r_p$.

{\bf Summary}---  In summary, this paper systematically explores the driven critical dynamics at the $1$D Ising tricrical point. We find that for driven dynamics along the critical line to cross the tricritical point, the KZ scaling characterized by exponent $r=z+1/\nu_\mu$ still stands even though the AIS breaks down in the critical region. We point out that the reason is that the excitations in the initial stage are much fewer compared to those at the critical point. In addition, we have also revealed that the driven dynamic along other direction is described by the KZ scaling with the typical exponent $r=z+1/\nu_p$, due to the dominating perpendicular relevant direction. This work provides profound new perspective of nonequilibrium critical dynamics near tricritical points. Recently, it was proposed that the Ising supersymmetric tricritical point can be realized in Rydberg atomic systems. Thus, it is expected that our results can be futher examined in these experiments~\cite{Keesling2019,king2023nature,Ebadi2021,garcia2024resolving,king2024computational,PhysRevB.106.L041109,PRXQuantum,science.abo6587}.

{\bf Note added}--- We thank the authors of~\cite{Htwang2025} for helpful discussion. In \cite{Htwang2025}, driven dynamics of the tricritical critical in another model belonging to the same universality class is considered and consistent results are obtained.

{\bf Acknowledgments}---  Y.-F.J. acknowledges support from National Key R$\&$D Program of China under Grants No. 2022YFA1402703 and from NSFC under Grant No. 12347107. S. Yin is supported by the National Natural Science Foundation of China (Grants No. 12222515 and No. 12075324) and the Science and Technology Projects in Guangdong Province (Grants No. 2021QN02X561) and Guangzhou City (Grant No. 2025A04J5408). 

\bibliographystyle{apsrev4-1}
\bibliography{ref}

% \appendix
\onecolumngrid
\newpage
\widetext
\thispagestyle{empty}

\setcounter{equation}{0}
\setcounter{figure}{0}
\setcounter{table}{0}
\renewcommand{\theequation}{S\arabic{equation}}
\renewcommand{\thefigure}{S\arabic{figure}}
\renewcommand{\thetable}{S\arabic{table}}

\pdfbookmark[0]{Supplementary Materials}{SM}
\begin{center}
    \vspace{3em}
    {\Large\textbf{Supplementary Materials for ``Driven Critical Dynamics in Tricitical Point"}}\\
%    \vspace{1em}
%    {}\\
    % \vspace{1.5em}
%    \vspace{0.5em}
\end{center}

\section{Determination of the tricritical point}
\begin{figure}[tbh]
\centering
  \includegraphics[width=0.66\linewidth,clip]{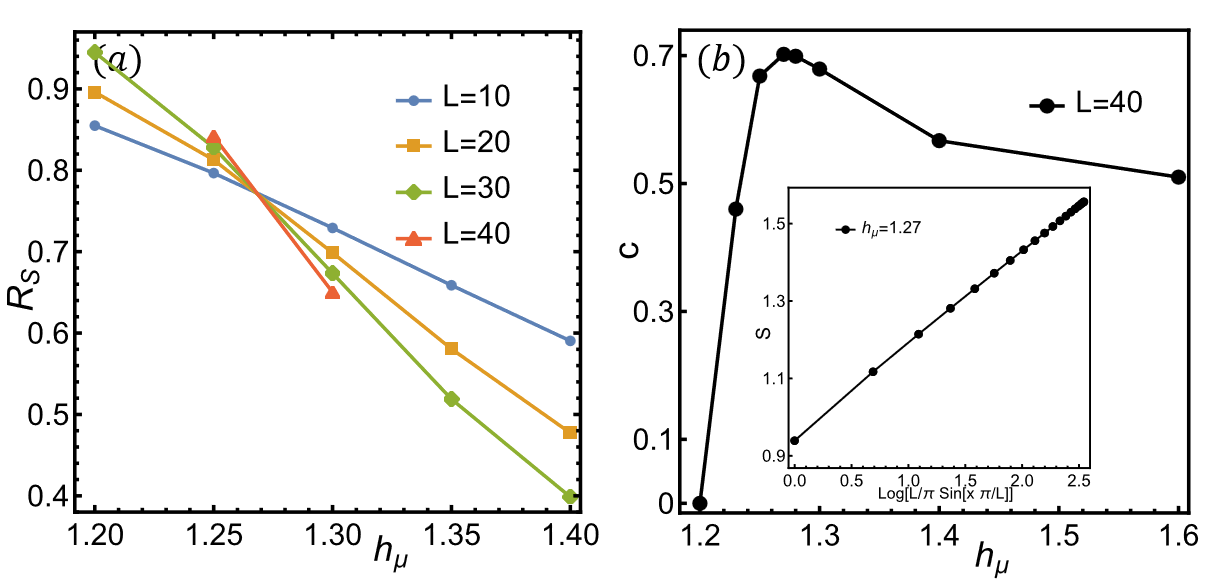}
  \vskip-3mm
  \caption{(a) The tricritical point $h_{\mu,c}$ determined by the crossing of the correlation ratio $R_S$ obtained from DMRG calculation on a series of chains with $L=10\sim 40$. (b) The central charge $c$ extracted from the entanglement entropy $S$ for systems with varies of $h_\mu=1.2\sim 1.6$ and fixed $h_\sigma =1$. Inset: the entanglement entropy $S(x)$ for the system at tricritical point $h_\mu \approx 1.27$. }
  \label{Sfigure1}
\end{figure}

To explore the critical dynamics near the tricritical point, we at first determine the critical point. Firstly, at the tricritical point, $h_\sigma$ should be zero, since the tricritical point is the terminal of the critical point. Second, to determine the value of $h_{\mu,c}$, we calculate the correlation ratio defined as:
\begin{equation}
	R_{S}\equiv{1-S(\bm \pi - \Delta \bm q)}/{S(\bm \pi)},
	\label{eq:RS}
\end{equation}
where $\Delta\bm{q}$ is the minimal lattice momentum and $S({\bm q})$ is the structure factor defined as
\begin{equation}
	S({\bm{q}})=  \frac{1}{L^{2d}} \sum_{i,j} \mathrm{e}^{\mathrm{i} \bm{q}\cdot ({\bm{r}_i-\bm{r}_j})} \langle{\mu_i^z \mu_j^z}\rangle,
	\label{eq:S_AFM}
\end{equation}
$R_{S}$ is a dimensionless quantity. Therefore, for fixed $g=0.5$ and $h_\sigma=1$, the curves of $R_{S}$ versus $h_\mu$ for different $L$ can cross at the tricritical point $h_{\mu,c}$. Fig.~\ref{Sfigure1}(a) shows the numerical results obtained via density-matrix renormalization group, where we can find that $h_{\mu,c}\approx 1.27$.

This value of the tricritical point can further be verified by the value of the central charge $c$ of the entanglement entropy extracted from
\begin{equation}
	S(x) = \frac{c}{3} \ln(\frac{L}{\pi} \sin[\frac{\pi x}{L}]) + const. .
	\label{eq:EE}
\end{equation}
Fig.~\ref{Sfigure1}(b) shows the value of the central charge versus $h_\mu$ for fixed $h_\sigma=1$. We finds that for $h_\mu<1.27$, $c=0.5$ corresponding to the conformal field theory of the Ising criticality; while for $h_\mu$ close to $1.27$, $c\approx 0.7$ which is the corresponding value of the Ising tricriticality.

\section{scaling in the process}
In driven dynamics from the Ising critical phase, we argue that the half-chain fermion correlation should satisfy $C_f\propto L^{-1}R^{\eta_f/r_\mu}$ for large $R$, in which $C_f\propto L^{-1}$ comes from the memory effects of the initial state, while $C_f\propto R^{\eta_f/r_\mu}$ reflects the tricritical property.

\begin{figure}[tbh]
\centering
  \includegraphics[width=0.4\linewidth,clip]{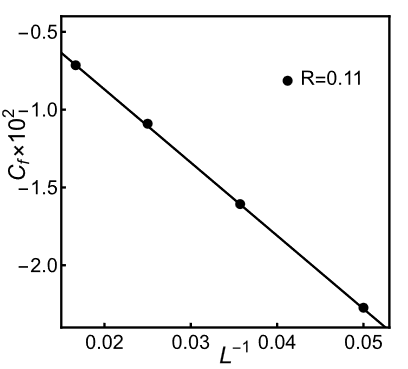}
  \vskip-3mm
  \caption{Scaling behavior of the half-chain fermion correlation $C_f$ at the tricritical point when the system is driven from the gapless Ising critical phase with large $R$. The solid line shows the scaling law $C_f\propto L^{-1}$ fitted from $L = 20 \sim 60$ data in large $R$ region.}
  \label{Sfigure2}
\end{figure}

In the main text, we demonstrate in Fig.~3 that $C_f\propto R^{\eta_f/r_\mu}$. Here we confirm that for large $R$, $C_f\propto L^{-1}$, as shown in Fig.~\ref{Sfigure2}.

\end{document}